# Poisoning effect of ammonia on the performance and transport process of proton exchange membrane fuel cells

Yaxian Han[1], Wei Gao[1], Yichao Huang[1], Tianyou Wang[1,2], Zhizhao Che[1,2,*]

1. State Key Laboratory of Engines, Tianjin University, Tianjin, 300350, China.

2. National Industry-Education Platform of Energy Storage, Tianjin University, Tianjin, 300350, China

*Corresponding author: chezhizhao@tju.edu.cn

# Abstract

Ammonia is a high-density hydrogen energy carrier and can be decomposed to produce hydrogen for use in fuel cells. However, a significant challenge in ammonia-decomposition-based fuel cell applications is the unavoidable presence of trace ammonia impurity, which can poison the fuel cell, but the poisoning mechanism remains unclear. To address this, a three-dimensional numerical model of proton exchange membrane (PEM) fuel cells with ammonia impurities is established to explore the transport process and underlying poisoning mechanism. The influences of key factors, including ammonia concentration, operating temperature, operating humidity, and membrane thickness, are studied. The poisoning mechanism is analyzed from the perspectives of the distributions of proton conductivity, current density, and dissolved water content. The results show that ammonia diminishes the cell performance by substantially reducing the proton conductivity of both the PEM and the anode catalyst layer. Higher operating temperatures and higher operating humidity can alleviate ammonia poisoning. Decreasing the membrane thickness can also help to mitigate ammonia poisoning, but may lead to less uniform current distribution.

# 1. Introduction

Proton exchange membrane fuel cell (PEMFC) is a clean and efficient hydrogen energy utilization device [1]. It converts chemical energy directly into electricity through the hydrogen oxidation reaction (HOR) at the anode and the oxygen reduction reaction (ORR) at the cathode, with water produced mainly at the cathode, and has the advantages of high efficiency, low noise, and zero emissions [2-6]. Ammonia ($NH_3$) is an important hydrogen energy carrier that has a high hydrogen storage density and is easy to transport and store [7-9]. It can be produced from hydrogen through ammonia synthesis processes, and decomposes to hydrogen and nitrogen when needed, thereby providing fuel for PEMFCs and becoming solutions for hydrogen storage and transportation [10-13]. However, during ammonia decomposition for hydrogen production, there will inevitably be ammonia impurities in the obtained hydrogen [14]. If ammonia enters a fuel cell with hydrogen, it can dissolve in water/ionomer phases and be protonated to form ammonium cation ($NH_4^+$), which can poison the fuel cell [15-17]. Therefore, hydrogen quality standards for PEMFC road-vehicle applications impose a stringent ammonia limit of 0.1 μmol/mol (0.1 ppm) [18,19]. Hence, it is critical to study the mechanism of ammonia poisoning in fuel cells to optimize the design, enhance the tolerance of fuel cells to ammonia, and promote the development of ammonia-hydrogen technology.

Previously, many scholars have studied the effect of $NH_3$ poisoning on PEMFC through a variety of experimental methods, and deduced the poisoning process and mechanism of $NH_3$ impurities through the experimental results. Hu et al. [20] studied the effects of different $NH_3$ concentrations in hydrogen on PEMFC and found that with an increase of $NH_3$ concentration in hydrogen, the voltage reduction became more obvious, and the reduction rate increased. Gomez et al. [21] experimentally found that 200 ppm of $NH_3$ in hydrogen rapidly reduced the output voltage of cells. Rajalakshmi et al. [22] found that when the $NH_3$ concentration exceeded 20 μL/L, $NH_3$ reacted with protons to generate $NH_4^+$, causing irreversible effects on the electrolyte membrane. The results of Uribe et al. [23] showed that $NH_3$ did not directly poison the anode or cathode catalyst, but reacted with protons in the electrolyte membrane to generate $NH_4^+$, which led to a decline in the conductivity of the electrolyte membrane. Halseid et al. [24] also reached a similar conclusion. Zhang et al. [25] used cyclic voltammetry and AC impedance testing to study the electrochemical performance of trace ammonia impurities in hydrogen in PEMFCs. The results showed that trace $NH_3$ impurities could seriously contaminate the membrane and catalyst layer, and affect proton transport and membrane conductivity by occupying charge sites. The influence of $NH_3$ impurities on the catalyst layer can be mostly recovered, while the influence on the membrane layer is not easy to recover. Soto et al. [26] studied the effect of transient ammonia concentration on PEMFC. Their results showed that the cell is able to fully recover from performance loss in neat $H_2$ after exposure to 200-ppm $NH_3$ for 10 hours, and there is a two-step process for the recovery. However, the detailed mechanism for the recovery is unclear.

From the perspective of the poisoning mechanism, it has been confirmed that ammonia poisoning is closely related to membrane conductivity. Hongsirikarn et al. [27] proposed that ammonium ion distribution critically influences Nafion membrane conductivity in operating PEMFCs. Hongsirikarn et al. [28] studied the impact of $NH_3$ and $NH_4^+$ poisoning on Nafion membrane conductivity by electrochemical impedance spectroscopy (EIS). They found that the resistance to ammonia poisoning could be enhanced by operating the PEMFC under high relative humidity, while the activation energy of proton conductivity exhibited positive correlation with ammonium concentration at particular humidity but negative dependence on relative humidity. These studies on the influence of ammonia on membrane conductivity provide a theoretical basis for analyzing the transport process and the influence of ammonia in fuel cells.

Besides cell-level observations, the influence of ammonium on perfluorosulfonic-acid (PFSA) ionomers has been investigated at the materials level. Halseid et al. [29] quantified the ion-exchange equilibrium and showed that increasing the ammonium concentration reduces both the water content and the proton conductivity of Nafion 117, with stronger penalties under gas-phase conditions. Lopes et al. [30] further measured ionic transport and water-vapor uptake in ammonium-exchanged PFSA membranes and reported reduced conductivity and altered sorption behavior. At the electrode level, ammonia/ammonium can also perturb electrocatalysis. An experimental study of hydrogen impurities by Wang et al. [31] summarized tolerance thresholds and characteristic voltage-loss behaviors for PEMFCs, including the strong sensitivity to $NH_3$. Electrochemical studies on polycrystalline platinum in perchloric acid showed that ammonium can affect ORR [32], and that trace ammonium ions can suppress ORR activity and increase peroxide formation on Pt/C catalysts, consistent with a site-blocking/oxidation pathway at high potentials [33]. Related measurements in acidic media by Halseid et al. [34] also reported changes in ORR kinetics in the presence of ammonium species. Fuel-cell-level studies by Yuan et al. [35] further discussed the impact of ammonia exposure at the cathode using in situ measurements including EIS and cyclic voltammetry (CV). Recent work by Jing et al. [36] reported pronounced impedance growth during ammonia exposure and quantified its evolution using equivalent circuit and distribution of relaxation times. Comparative tests in low- and high-temperature PEMFCs by Isorna et al. [37] also indicated that the $NH_3$ impact can vary with membrane chemistry and water management.

Mechanistic insight at the nanoscale has also advanced. Recent molecular-dynamics studies showed that ammonia/ammonium can interact strongly with PFSA side chains and water clusters, forming ion-water network structures that hinder proton diffusion at low hydration [38], and can replace hydronium at sulfonate sites and promote ammonia-derivative ion clustering in the cathode catalyst-layer ionomer, thereby blocking proton-transport pathways [17]. Complementary adsorption/desorption measurements demonstrated that $NH_3$ can persist in Nafion and exchange with protons, highlighting the importance of uptake and release processes [39]. Collectively, these findings suggest that ammonia poisoning arises from coupled cation exchange, hydration-dependent transport, and electrochemical feedbacks across multiple length scales. Very recent PEMFC studies have further linked ammonia exposure to experimentally observable diagnostics, including current-density non-uniformity with strong relative-humidity dependence under airborne ammonia contamination [40] and changes in key resistance/overpotential components under $NH_3$ impurities [41].

Previous studies on $NH_3$ contamination in PEMFCs have provided valuable experimental evidence and trend-level interpretations. However, existing simulations either adopt one-dimensional or lumped descriptions, or treat ammonia-related cation exchange via spatially uniform effective conductivity or ohmic-loss adjustments [15,42], which do not resolve the coupled transport fields and spatial localization of poisoning inside the membrane electrode assemblies (MEA). In particular, the coupled pathway from $NH_3$ transport and uptake into ionomer, to ammonium formation and sulfonic-site neutralization, and further to hydration-dependent proton conductivity, current redistribution under channel-rib geometry, and feedback to water transport has not been quantitatively established in a transport-resolved framework.

In this work, we develop a three-dimensional multiphysics model that explicitly couples $NH_3$ transport in the anode flow field and porous layers, its dissolution in ionomer phases, ammonium formation in hydrated ionomer and partial neutralization of sulfonic acid sites, and the resulting conductivity change in both the PEM and anode catalyst-layer (ACL) ionomer. This enables spatially resolved diagnostics (hydration, conductivity, and current density distributions) and mechanistic interpretation of the non-uniform performance loss along the channel and across under-rib/under-channel regions. Compared with previous effective or lumped ammonia-poisoning models, the present model does not prescribe $NH_3$ poisoning as a spatially uniform voltage loss, empirical ohmic-resistance increase, or uniform conductivity correction. Instead, $NH_3$ is treated as an explicitly transported species in the anode flow field and porous layers. Its dissolution into the hydrated ionomer phase, protonation to $NH_4^+$, partial neutralization of sulfonic acid sites, and the resulting decrease in proton conductivity are locally coupled with gas transport, water transport, charge transport, and electrochemical reactions. Compared with conventional 3D PEMFC models that mainly resolve multiphase transport and electrochemical processes without ammonia-specific ionomer chemistry, the present model further introduces $NH_3/NH_4^+$-induced ionomer-site neutralization in both the PEM and the ACL ionomer phase. Therefore, the model can quantify not only the overall polarization and power-density loss, but also the spatial distributions of dissolved water content, proton conductivity, and local current density.

# 2. Mathematical model

## 2.1 Geometric model

In this study, the structure of the PEMFC mainly includes a bipolar plate, gas flow channels (GFC), gas diffusion layers (GDL), catalyst layers (CL), microporous layers (MPL), and a proton exchange membrane (PEM). The model structure is presented in Figure 1. It consists of one channel and one adjacent rib with channel width 1 mm, channel height 1.2 mm, and rib width 1 mm; and the channel length is 50 mm. The thicknesses of the GDL, CL, MPL, and PEM are 0.26 mm, 0.01 mm, 0.02 mm, and 50.8 μm, respectively (unless otherwise stated). The key parameters are summarized in Table S1. The 3D computational domain, consisting of one channel and one adjacent rib, represents a transverse repeating unit of the flow-field. This channel-rib repeating unit approach is widely used in PEMFC multiphysics modeling because the channel pattern is periodic in the transverse direction for typical parallel-channel flow fields. The present domain can capture not only the inlet-to-outlet gradients along the channel, but also the non-uniformity between the under-rib and under-channel regions across the MEA, which is essential for quantifying spatial localization of hydration, conductivity, and current density under $NH_3$ poisoning. In addition, since the channel has a large streamwise length-to-width ratio, the streamwise direction is displayed with a shortened aspect ratio (1/4 of the true length) in all figures presented in this study to improve clarity; however, the numerical simulations are performed on the full channel length.

The mathematical model incorporates these fundamental assumptions: (1) The flow and transport process are in steady-state; (2) The gas in the GFC remains laminar and follows the ideal gas law; (3) The porous media in the CL, MPL, and GDL are isotropic and uniform; (4) The influence of gravity is neglected; (5) Reactant-gas crossover through the PEM is neglected in the present model to focus on the dominant coupled transport/electrochemical processes associated with $NH_3$-induced proton-transport degradation. Gas crossover can occur in practical cells via diffusion, and may slightly affect local partial pressures and water balance, particularly at high current density.

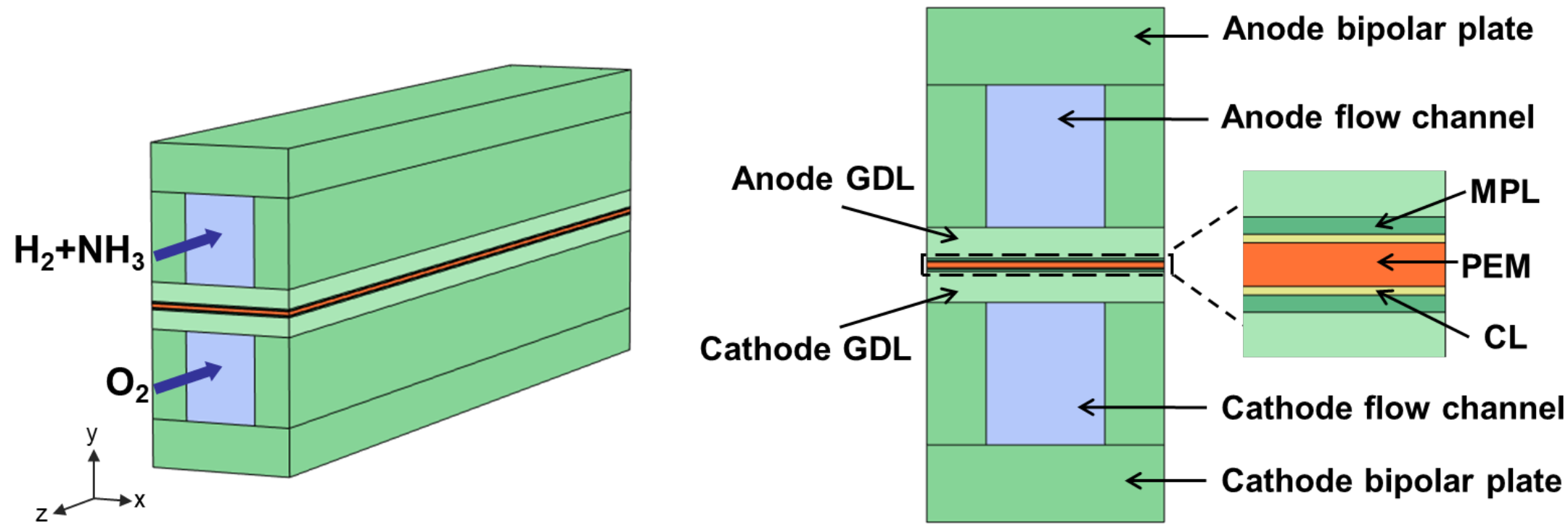

Figure 1. Schematic illustration of the geometric model of the PEMFC (not to scale).

## 2.2 Multiphysics modeling

To describe the flow, phase change, reactant transport, and electrochemical reaction in a PEMFC supplied with $NH_3$-contaminated hydrogen impurities, a multiphysics numerical model of PEMFC with $NH_3$ impurities was established in this study, which mainly considers the conservation and transport of mass, momentum, energy, species, water, and charges. The governing equations implemented in the present computational model comprise:

Mass conservation equation for gas species (GFC, GDL, MPL, and CL):

$$\nabla \cdot \left( \rho_{\mathrm{g}} \mathbf{u}_{\mathrm{g}} \right) = S_{\mathrm{m}} \tag{1}$$

Momentum transport equation governing gas dynamics (GFC, GDL, MPL, and CL):

$$\nabla \cdot \left( \frac{\rho_{\mathrm{g}} \mathbf{u}_{\mathrm{g}} \mathbf{u}_{\mathrm{g}}}{\varepsilon^2 (1-s)^2} \right) = -\nabla P_{\mathrm{g}} + \mu_{\mathrm{g}} \nabla \cdot \left[ \nabla \left( \frac{\mathbf{u}_{\mathrm{g}}}{\varepsilon (1-s)} \right) + \nabla \left( \frac{\mathbf{u}_{\mathrm{g}}^{\mathrm{T}}}{\varepsilon (1-s)} \right) \right] - \frac{2}{3} \mu_{\mathrm{g}} \nabla \left[ \nabla \cdot \left( \frac{\mathbf{u}_{\mathrm{g}}}{\varepsilon (1-s)} \right) \right] + \mathbf{S}_{\mathrm{u}} \tag{2}$$

Gas species transport equation (GFC, GDL, MPL, and CL):

$$\nabla \cdot \left( \rho_{\mathrm{g}} \mathbf{u}_{\mathrm{g}} Y_{\mathrm{i}} \right) = \nabla \cdot \left( \rho_{\mathrm{g}} D_{\mathrm{i}}^{\mathrm{eff}} \nabla Y_{\mathrm{i}} \right) + S_{\mathrm{i}} \tag{3}$$

where the subscript “i” stands for hydrogen, ammonia, oxygen, water vapor, and nitrogen. The diffusive mass flux of gas species is described by Fick’s law $\mathbf{j}_{\mathrm{i}} = -\rho_{\mathrm{g}} D_{\mathrm{i}}^{\mathrm{eff}} \nabla Y_{\mathrm{i}}$, where $\mathbf{j}_{\mathrm{i}}$ is the diffusive mass flux, $\rho_{\mathrm{g}}$ is the gas-mixture density, $D_{\mathrm{i}}^{\mathrm{eff}}$ is the effective diffusion coefficient, and $Y_{\mathrm{i}}$ is the mass fraction of species i. In the gas flow channels, $D_{\mathrm{i}}^{\mathrm{eff}}$ is the gas-phase diffusion coefficient, while in the porous layers it is corrected by the porosity and liquid-water saturation.

Liquid water saturation transport equation in porous media (GDL, MPL, and CL):

$$\nabla \cdot \left( \frac{\rho_{\mathrm{l}} K_{\mathrm{l}}}{\mu_{\mathrm{l}}} \nabla (p_{\mathrm{g}} - p_{\mathrm{c}}) \right) + S_{\mathrm{l}} = 0 \tag{4}$$

The liquid-water saturation equation is applied only in the porous layers, including the GDL, MPL, and CL. In these porous layers, liquid-water transport is governed by capillary-pressure-driven flow and relative permeability. The GFC is treated as a free-flow gas domain, and the liquid-water motion in the GFC is not explicitly resolved.

Dissolved water conservation equation (CL and PEM):

$$\nabla \cdot \left( n_{\mathrm{d}} \frac{\mathbf{J}_{\mathrm{ion}}}{F} \right) = \frac{\rho_{\mathrm{mem}}}{EW} \nabla \cdot \left( D_{\mathrm{d}}^{\mathrm{eff}} \nabla \lambda \right) + S_{\mathrm{d}} \tag{5}$$

Electron charge transport equation (GDL, MPL, BP, and CL):

$$\nabla \cdot \left( \sigma_{\mathrm{e}}^{\mathrm{eff}} \nabla \varphi_{\mathrm{e}} \right) + S_{\mathrm{e}} = 0 \tag{6}$$

Ionic charge transport equation (CL and PEM):

$$\nabla \cdot \left( \sigma_{\mathrm{ion}}^{\mathrm{eff}} \nabla \varphi_{\mathrm{ion}} \right) + S_{\mathrm{ion}} = 0 \tag{7}$$

Energy conservation equation (GFC, GDL, MPL, BP, CL, and PEM):

$$\nabla \cdot \left( \left( \rho C_{\mathrm{p}} \right)_{\mathrm{fl}}^{\mathrm{eff}} \mathbf{u} T \right) = \nabla \cdot \left( k_{\mathrm{fl,sl}}^{\mathrm{eff}} \nabla T \right) + S_{\mathrm{T}} \tag{8}$$

The corresponding parameters and the source terms used in the model are listed in Tables S2 and S3 in the Supplementary Material, respectively.

## 2.3 Mechanism of ammonia poisoning

$NH_3$ is treated as an explicit species in the anode stream and porous media. It is transported by convection/diffusion in the anode channel and by effective diffusion in porous layers, and a fraction dissolves into the ionomer phase. In hydrated ionomer, dissolved $NH_3$ is protonated to form ammonium ions ($NH_4^+$), which partially neutralize sulfonic acid sites. The equilibrium relations are as follows:

$$\mathrm{NH_3(g)} \leftrightarrow \mathrm{NH_3(mem)} \tag{9}$$

$$\mathrm{NH_3(mem)} + \mathrm{H^+} \leftrightarrow \mathrm{NH_4^+} \tag{10}$$

The presence of ammonium significantly affects the proton conduction capability of the electrolyte in the PEM and ACL. According to the model established by Pisani et al. [43] and Hongsirikarn et al. [28], when $2 < \lambda < 10$, the membrane conductivity is:

$$\sigma = \left( \frac{\lambda}{\frac{EV}{18} + \lambda} - \varepsilon_0 \right)^t \left[ c_{\mathrm{H+}} \left( K_V \left( 1 - \frac{n_{\mathrm{sub}}}{\lambda} \right) + K_S \left( 1 - \frac{n_{\mathrm{sub}}}{\lambda} \right) - \left( \frac{\lambda^4}{\left( n_0^s \right)^4} + 1 \right)^{-1/4} \right) \exp\left( -\frac{n_{\mathrm{pol+}}}{\lambda - n_{\mathrm{sub}}} \right) + \frac{\frac{1}{32} c_{\mathrm{H_20}}^2 F^2 d_0^2}{\sqrt{\mu^2 / \left( n_0^s \right)^4 + \mu^2 / (\pi \lambda)^4}} \right] \exp\left( -\frac{E_{\mathrm{a}}}{RT} \right) \tag{11}$$

$$E_{\mathrm{a}} = 12.46 + 5.87 y_{\mathrm{NH_4^+}} - 5.20 RH_{\mathrm{a}} \tag{12}$$

where $E_a$ represents the activation energy of proton conductivity, kJ/mol, $y_{\mathrm{NH_4^+}}$ is the ratio of the sulfonic acid site neutralized by the ammonium ion to the initial total sulfonic acid site.

For the ACL, the proton conductivity of the ionomer phase is also calculated according to Eq. (11). In this way, the $NH_3$-induced neutralization effect is applied to the proton conductivity of the PEM and the ionomer phase in the ACL. Because ammonia crossover to the cathode side is neglected, $y_{\mathrm{NH_4^+}}$ on the cathode side is 0. The values of the variables involved in Eqs. (11)-(12) are listed in Table 1.

The $NH_3$-induced degradation in the ACL considered in the present model refers mainly to proton-transport degradation. Previous studies have shown that ammonia poisoning in PEMFCs with ammonia-containing hydrogen is strongly associated with the formation of $NH_4^+$, ionomer-site neutralization, and the resulting decrease in membrane/ionomer proton conductivity [23,24,27-29]. This mechanism provides a direct basis for introducing the $NH_3/NH_4^+$-induced conductivity loss in both the PEM and the ACL ionomer phase. It should be noted that ammonia/ammonium may also affect Pt electrocatalysis through adsorbed nitrogen-containing intermediates under certain electrochemical conditions [32,44]. This pathway is not explicitly included here because its contribution to anode HOR loss is condition-dependent and requires additional adsorption/oxidation kinetics and surface-coverage parameters. Since these kinetic parameters are not available, the current framework focuses only on the conductivity-dominated poisoning pathway.

Table 1. Parameters for the conductivity model [43].

| Parameters | Value |
|---|---|
| $K_V$ | 8 S m$^{-1}$ |
| $K_S$ | 25.1 S m$^{-1}$ |
| $n_{sub}$ | 2 |
| $n_{pol+}$ | 5 |
| $\mu$ | 8.9×10$^{-4}$ kg m$^{-1}$ s$^{-1}$ |
| $n_0^s$ | 12 |
| $F$ | 96485 C mol$^{-1}$ |
| $d_0$ | 1.2×10$^{-7}$ m |
| $t$ | 1.75 |
| $EW$ | 1100 kg kmol$^{-1}$ |
| $\rho_{mem}$ | 1980 kg m$^{-3}$ |
| $\varepsilon_0$ | 0 |

The coupling pathway for $NH_3$ poisoning in the present 3D multiphysics model is schematically summarized in Figure 2. The resulting conductivity change modifies the local ohmic loss and redistributes the current density under the channel-rib geometry. This current redistribution feeds back to local water production and water transport (electroosmotic drag and back diffusion), thereby altering the hydration level, which in turn further affects the conductivity and local reaction rates. Through this strongly coupled loop, the model predicts spatially non-uniform performance loss along the flow direction and across under-rib/under-channel regions.

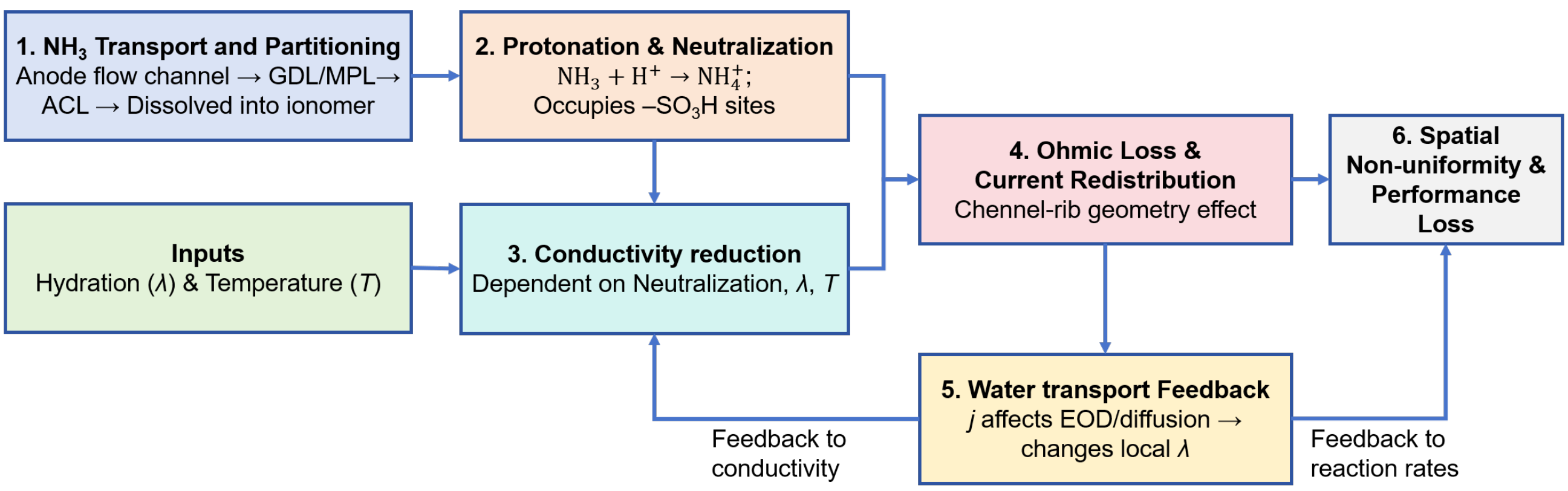

Figure 2. Schematic of the coupling pathway for $NH_3$ poisoning in the multiphysics model.

Compared with lumped or 1D treatments that apply an empirical voltage penalty or a spatially uniform conductivity reduction, the present 3D framework resolves (1) $NH_3$ transport and uptake, (2) neutralization-dependent conductivity in both PEM and ACL ionomer, and (3) the resulting current-density redistribution and its feedback to hydration. This enables internal-field-based interpretation and geometry-dependent non-uniformity assessment that are not accessible in lumped approaches.

This work focuses on short-to-medium timescale performance losses induced by $NH_3$ through ionomer-site neutralization and the resulting proton-transport disruption in the PEM and ACL ionomer. Potential long-term material-compatibility effects (e.g., corrosion/coating degradation or contact-resistance evolution in bipolar plates and GDL) are not explicitly modeled because they require time-dependent sub-models and dedicated parameter identification.

The range of $NH_3$ concentration investigated in this study (0-150 ppm) is selected to balance practical relevance and mechanistic sensitivity. From a standards perspective, hydrogen quality standards for PEMFC applications impose a stringent $NH_3$ limit on the order of 0.1 μmol/mol (i.e., 0.1 ppm) [18,19], indicating that even trace $NH_3$ can be critical for real deployment. From a research perspective, ppm-level $NH_3$ is frequently adopted in experimental studies to produce measurable and repeatable performance changes and to characterize sensitivity, reversibility, and the dependence on operating conditions. Therefore, the concentrations considered here are consistent with the experimental conditions used for validation and with recent $NH_3$ contamination literature, enabling systematic parametric analysis while retaining clear relevance to practical fuel-quality requirements.

## 2.4 Boundary conditions

The inlet mass flow rates are calculated at specified current density using hydrogen and oxygen stoichiometric ratios:

$$\dot{m}_{\mathrm{in,a}} = \frac{\rho_{\mathrm{g}}^{\mathrm{a}} \xi^{\mathrm{a}} I_{\mathrm{ref}} A_{\mathrm{mem}}}{2FC_{\mathrm{H_2}}} \tag{13}$$

$$\dot{m}_{\mathrm{in,c}} = \frac{\rho_{\mathrm{g}}^{c} \xi^{\mathrm{c}} I_{\mathrm{ref}} A_{\mathrm{mem}}}{4FC_{\mathrm{O_2}}} \tag{14}$$

The inlet concentrations of hydrogen (anode) and oxygen (cathode) are given by:

$$C_{\mathrm{H_2}} = \frac{\left(P_{\mathrm{g}}^{\mathrm{a}} - RH_{\mathrm{a}} P^{\mathrm{sat}}\right)}{RT_0} \tag{15}$$

$$C_{\mathrm{O_2}} = \frac{0.21\left(P_{\mathrm{g}}^{\mathrm{c}} - RH_{\mathrm{c}} P^{\mathrm{sat}}\right)}{RT_0} \tag{16}$$

where $T_0$ is the inlet temperature. The electronic potential on the anode plate surface is defined as 0 V, and the potential at the cathode plate boundary is defined as the difference between the fuel cell's reversible voltage and its operating voltage. The boundary potential for bipolar plates is:

$$\begin{cases} \varphi_{\mathrm{e}}^{\mathrm{a,end}} = 0 \\ \varphi_{\mathrm{e}}^{\mathrm{c,end}} = V_{\mathrm{cell}} \end{cases} \tag{17}$$

For the thermal boundary conditions, the external wall surfaces of the bipolar plates were set as isothermal boundaries at the prescribed operating temperature:

$$T = T_{\mathrm{wall}} = T_0 \tag{18}$$

where $T_{\mathrm{wall}}$ is the wall temperature and $T_0$ is the operating temperature. The inlet gas temperature was set to the same operating temperature:

$$T_{\mathrm{in}} = T_0 \tag{19}$$

The gas outlets were treated with convective outflow thermal boundary conditions. The lateral symmetry boundaries were treated as thermally symmetric/adiabatic boundaries:

$$-\mathbf{n} \cdot k_{\mathrm{eff}} \nabla T = 0 \tag{20}$$

To represent the periodic repetition of channels in the transverse direction, symmetry boundary conditions are applied on the two lateral faces of the computational domain, such that the solution corresponds to an infinitely repeating channel-rib pattern. In this way, the modeled unit is representative of the interior region of a multi-channel cell.

## 2.5 Model validation

The model validation and mesh-independence study are provided in the Supplementary Material. Briefly, the numerical model was validated against the experimental polarization curves reported by Uribe et al. [23] under 0 ppm and 30 ppm $NH_3$ conditions. As shown in Fig. S1, the simulated polarization curves agree reasonably well with the experimental data, with maximum absolute voltage errors of 0.0365 V at 0 ppm $NH_3$ and 0.0521 V at 30 ppm $NH_3$. The mesh-independence study also shows that the selected mesh (with 99,200 cells) provides sufficient numerical accuracy, with a maximum difference in limiting current density of less than 0.5% relative to the densest mesh.

# 3. Results and analysis

## 3.1 Influence of ammonia concentration

Figure 3 illustrates the influence of varying ammonia concentrations in hydrogen on cell performance. Figure 3(a) presents the polarization curve of PEMFC with varying $NH_3$ concentrations. Under identical working conditions, the maximum current density and peak power density decrease sharply with increasing $NH_3$ concentration. When the ammonia concentration increases to 100 ppm, the peak power density decreases by 21.1% under the steady-state poisoning condition considered in this model. This result indicates that ppm-level $NH_3$ contamination can cause significant performance loss for $NH_3$ uptake, $NH_4^+$ formation, and ionomer-site neutralization to develop. It should be noted that the polarization curves in Figure 3 are steady-state simulation results. They represent the self-consistent cell performance after $NH_3$ transport, dissolution into the ionomer phase, protonation to $NH_4^+$, sulfonic-site neutralization, and the resulting proton-conductivity degradation have been established for each prescribed $NH_3$ concentration. Therefore, the curves do not describe the short-term transient response immediately after $NH_3$ is introduced. They also do not represent long-term irreversible material degradation. Instead, they characterize the steady-state or quasi-equilibrated poisoning effect associated with $NH_3/NH_4^+$-induced ionomer proton-conductivity loss.

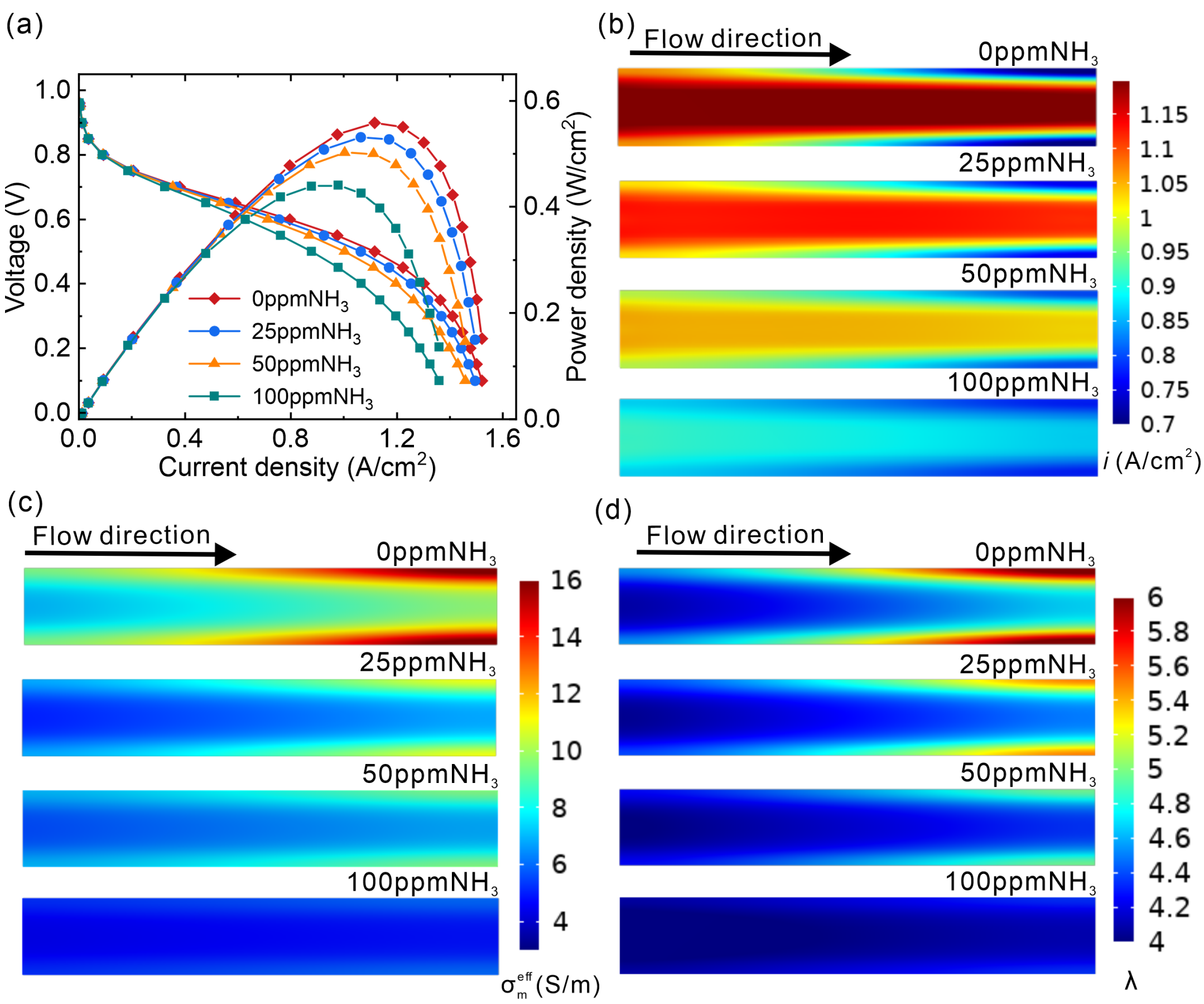

Figure 3. Influence of $NH_3$ concentrations on PEMFCs: (a) polarization and power-density curves; (b) current density distribution in the membrane central cross-section; (c) proton conductivity distribution at the membrane central cross-section; (d) distribution of dissolved water in the membrane central cross-section. The output voltage in (b-d) is 0.5 V. For visualization, the streamwise axes in panels (b-d) are displayed with a shortened aspect ratio (1/4 of the true length).

To further analyze the transport process in PEMFCs caused by ammonia gas, the current density distribution under different ammonia concentrations is compared and analyzed. Figure 3(b) reveals that the current density on the entire section has a substantial decrease. The decreasing current density can be further analyzed from the proton conductivity and dissolved water distribution in the membrane. Figure 3(c) suggests that the membrane conductivity $\sigma_{\mathrm{ion}}^{\mathrm{eff}}$ has a monotonic decrease as the $NH_3$ concentration increases. Meanwhile, the dissolved water content decreases gradually, which is consistent with the trend of the proton conductivity. This behavior can be attributed to the coupled effect of $NH_3/NH_4^+$-induced ionomer-site neutralization and water-transport redistribution. After $NH_3$ enters the cell with the hydrogen stream, it dissolves into the hydrated ionomer phase and is protonated to form $NH_4^+$. The formed $NH_4^+$ partially neutralizes sulfonic acid sites, reducing the number of effective proton-conducting sites and decreasing the local ionomer conductivity. The reduced current density then changes electroosmotic drag and water production, leading to a redistribution of dissolved water within the membrane. Since ionomer proton conductivity is strongly hydration-dependent, this altered hydration field further feeds back to the local proton conductivity. As the $NH_3$ concentration increases, the neutralization effect becomes stronger, the coupled conductivity-hydration degradation becomes more pronounced, and the cell performance decreases accordingly. This conductivity-driven increase in resistive losses is consistent with experimental observations showing increased resistance/overpotential contributions and impedance signatures under $NH_3$ exposure [36,41].

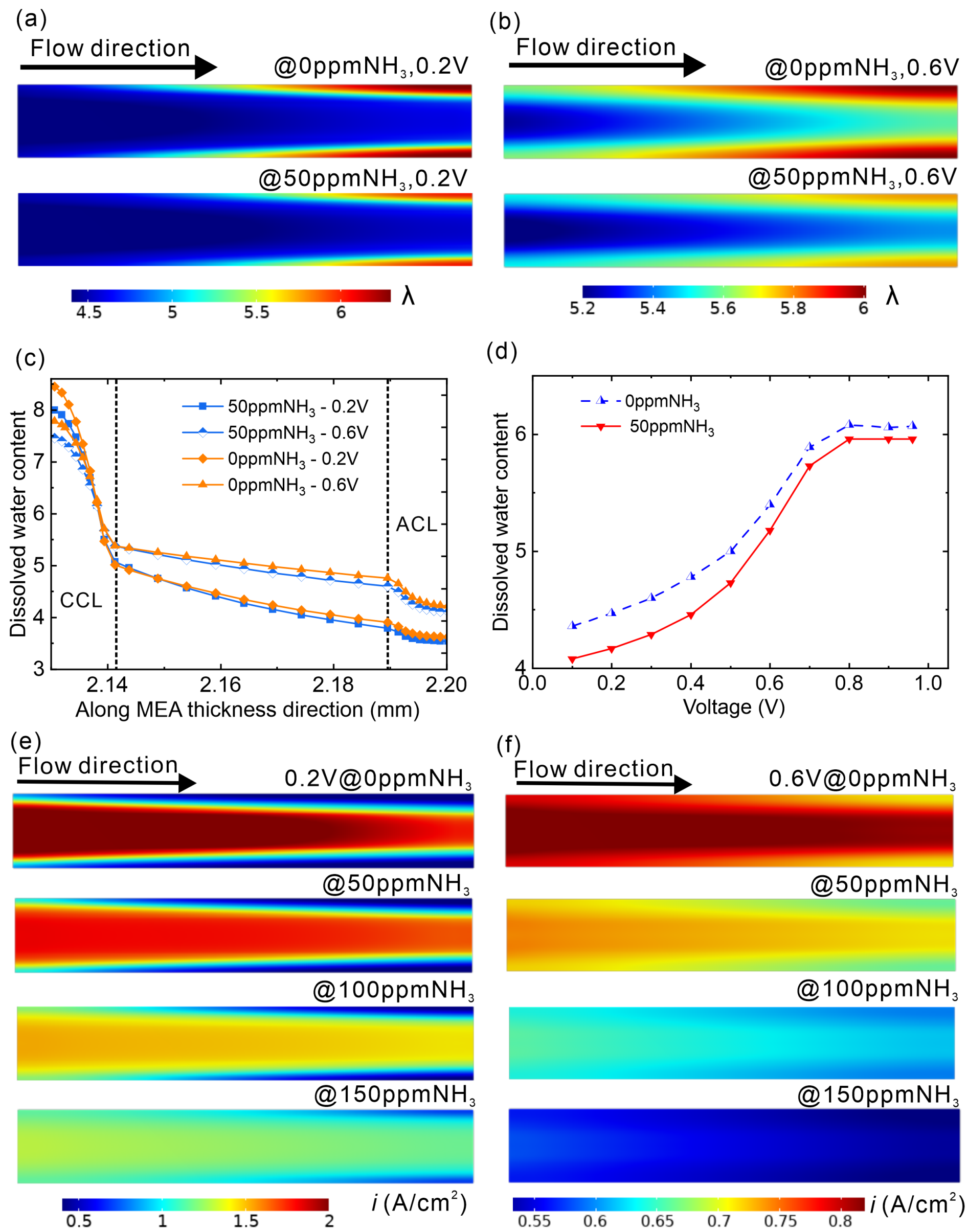

Figure 4. (a-b) Comparison of the distribution of the dissolved water content for 0 and 50 ppm $NH_3$ across the membrane's central cross-section at different voltages: (a) 0.2 V, (b) 0.6 V. (c) Dissolved water content distribution across the membrane thickness. (d) Average dissolved water content of 0 and 50 ppm $NH_3$ at the ACL-PEM interface under varying operating voltages. (e-f) Distribution of local current density at the central cross-section of the PEM at different operating voltages and $NH_3$ concentrations. For visualization, the streamwise axes in panels (a-b,e-f) are displayed with a shortened aspect ratio (1/4 of the true length).

It should be noted that the internal fields in Figure 3(b-d) are compared at a fixed cell voltage (e.g., 0.5 V). This is intended to represent the self-consistent response of the coupled PEMFC system under potentiostatic operation. Under $NH_3$ contamination, the current density at a given voltage changes and, consequently, the water production and hydration state also change; therefore, fixed-voltage comparisons reflect the combined effects of electrochemistry, mass transport, and ionomer hydration.

In view of the importance of the dissolved water content on proton conductivity, we analyze the dissolved water distribution in the membrane at different ammonia concentrations, as shown in Figure 4. Owing to water transport through flow channels, the dissolved water content of downstream regions exhibits elevated hydration levels compared to upstream, and the dissolved water content under the channel is lower than rib-covered zones. At both 0.6 and 0.2 V, the dissolved water content inside the cell with $NH_3$ impurities is lower than without $NH_3$ impurities, as shown in Figures 4(a) and 4(b). Compared with pure hydrogen, the current density at 50 ppm ammonia decreases with increasing cell potential, which weakens the electroosmotic transport of dissolved water from anode to cathode, resulting in an increase of dissolved water in the ACL and PEM. Meanwhile, due to the declining electrochemical reaction rate, the water content generated in the cathode catalyst layer (CCL) decreases, thus reducing the dissolved water content in the CCL ionomer, as demonstrated in Figure 4(c). Furthermore, the mean dissolved water content at the ACL-PEM interface under different operating voltages is shown in Figure 4(d). At lower operating voltages, the dissolved water content at 50 ppm $NH_3$ decreases relatively faster than that of pure hydrogen. Through the action of dissolved water content and proton conductivity, with increasing $NH_3$ concentration at both high and low operating voltages, the cell performance is reduced.

In addition to affecting the PEM, ammonia also affects the ACL. The ACL is composed of catalyst, carbon support, and ionomer, and its conductivity includes electron conductivity and proton conductivity. After the proton is generated in the ACL, it is transported to the PEM through the ionomer and ultimately reaches the cathode to participate in reactions. Therefore, ammonia impurity in the hydrogen gas will also affect the proton conductivity of the ACL. As evidenced in Figure 5, with increasing ammonia concentration, the proton conductivity gradually decreases. This is because ammonia inhibits the transport of protons in the ACL, and thus reduces the proton conductivity of the ACL. As ammonia concentration increases from 0 to 150 ppm, the average proton conductivity of the ACL decreases by 68.4%, which severely degrades cell performance. This is consistent with the polarization trend presented in Figure 3(a).

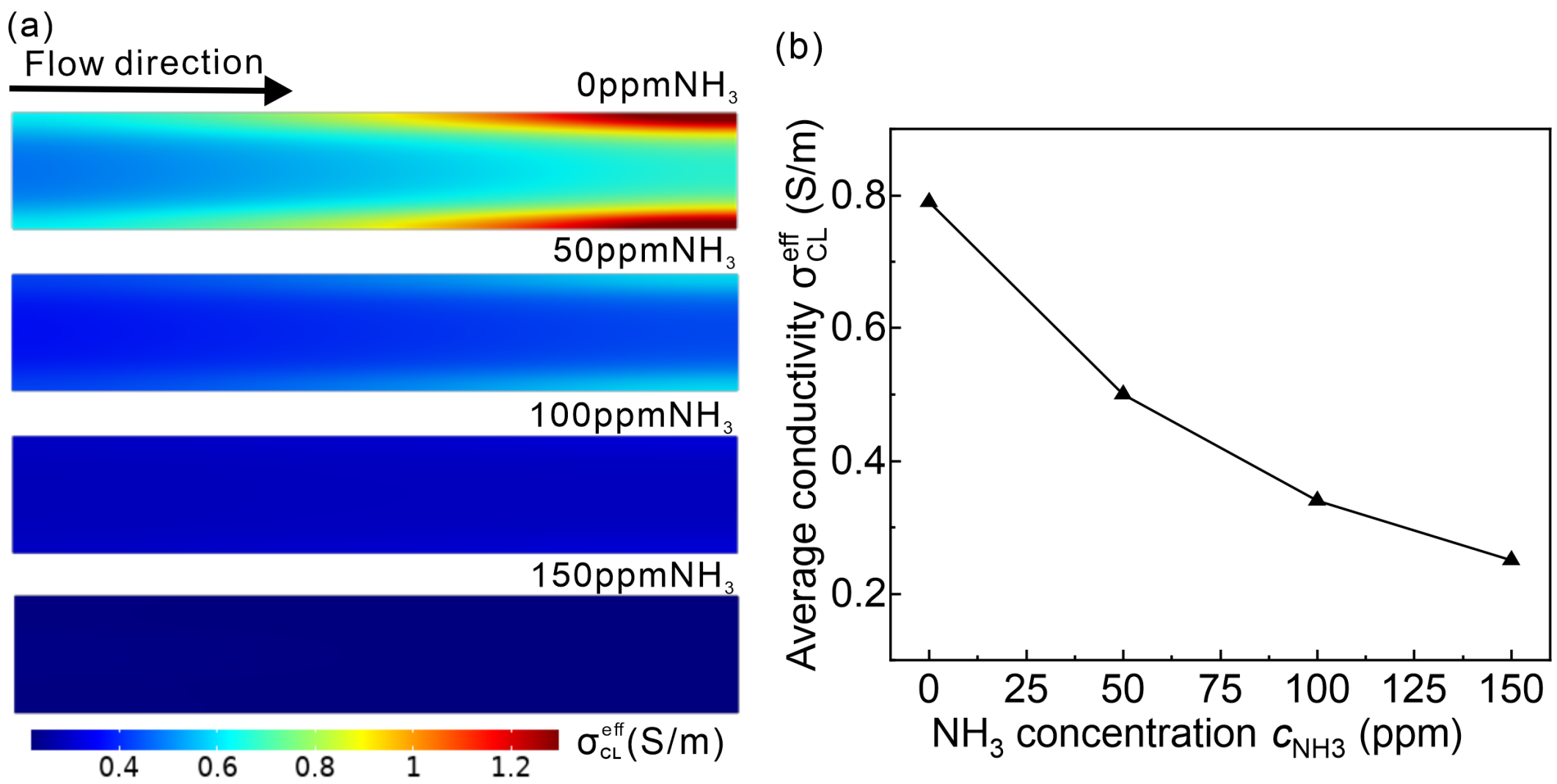

Figure 5. Influence of $NH_3$ concentration on the ACL: (a) proton conductivity distribution at the center cross-section of the ACL; (b) average proton conductivity of the ACL. For visualization, the streamwise axes in panel (a) are displayed with a shortened aspect ratio (1/4 of the true length).

## 3.2 Influence of operating temperature

Operating temperature serves as a critical parameter governing transport process and performance in PEMFCs. The impact of $NH_3$ impurities on fuel cells is studied at different temperatures. Figure 6 suggests that as operating temperature increases, the cell performance reduction when introducing the same amount of ammonia becomes smaller. At 60 °C, as the $NH_3$ concentration rises from 0 to 150 ppm, the peak power density decreases by 55.4%; while at 90 °C, the peak power density is reduced by 13.4%. Increasing temperature greatly mitigates the performance reduction by $NH_3$. This is because increasing temperature enhances reaction kinetics and transport properties, including gas diffusivity and ionomer proton mobility at a given hydration level, thereby reducing the apparent performance loss caused by $NH_3$-induced conductivity degradation. This also indicates that increasing temperature can enhance the PEMFC performance and tolerance to $NH_3$. The analysis of the membrane conductivity shows that the increase in temperature can improve the membrane conductivity of poisoned PEMFCs and enhance the tolerance of PEMFCs to $NH_3$. The influence of temperature on the ACL proton conductivity is consistent with the PEM, i.e., increasing the operating temperature is also conducive to mitigating the $NH_3$ poisoning effect on the ACL. The details of the influence of the operating temperature on the conductivities of PEM and ACL are provided in Section S2 of the Supplementary Material, including the distribution of proton conductivities in the PEM and ACL (Figures S2 and S3). These findings align with the molecular dynamics simulations reported by Huang et al. [17], which show that elevated temperature can lead to the dissociation of large ion clusters, mitigate the clogging effect of ammonium ions in the ionomer, and restore proton transport efficiency to a certain extent.

The present temperature analysis is performed within a typical operating window of low-temperature PEMFC (60-90 °C). Within this range, increasing temperature generally improves reaction kinetics and enhances transport (e.g., higher diffusivity and higher proton mobility at a given hydration level), which alleviates the apparent performance loss under $NH_3$ contamination. However, excessively high temperature may lead to membrane dehydration if inlet humidification and water balance are not adjusted, increasing ohmic loss.

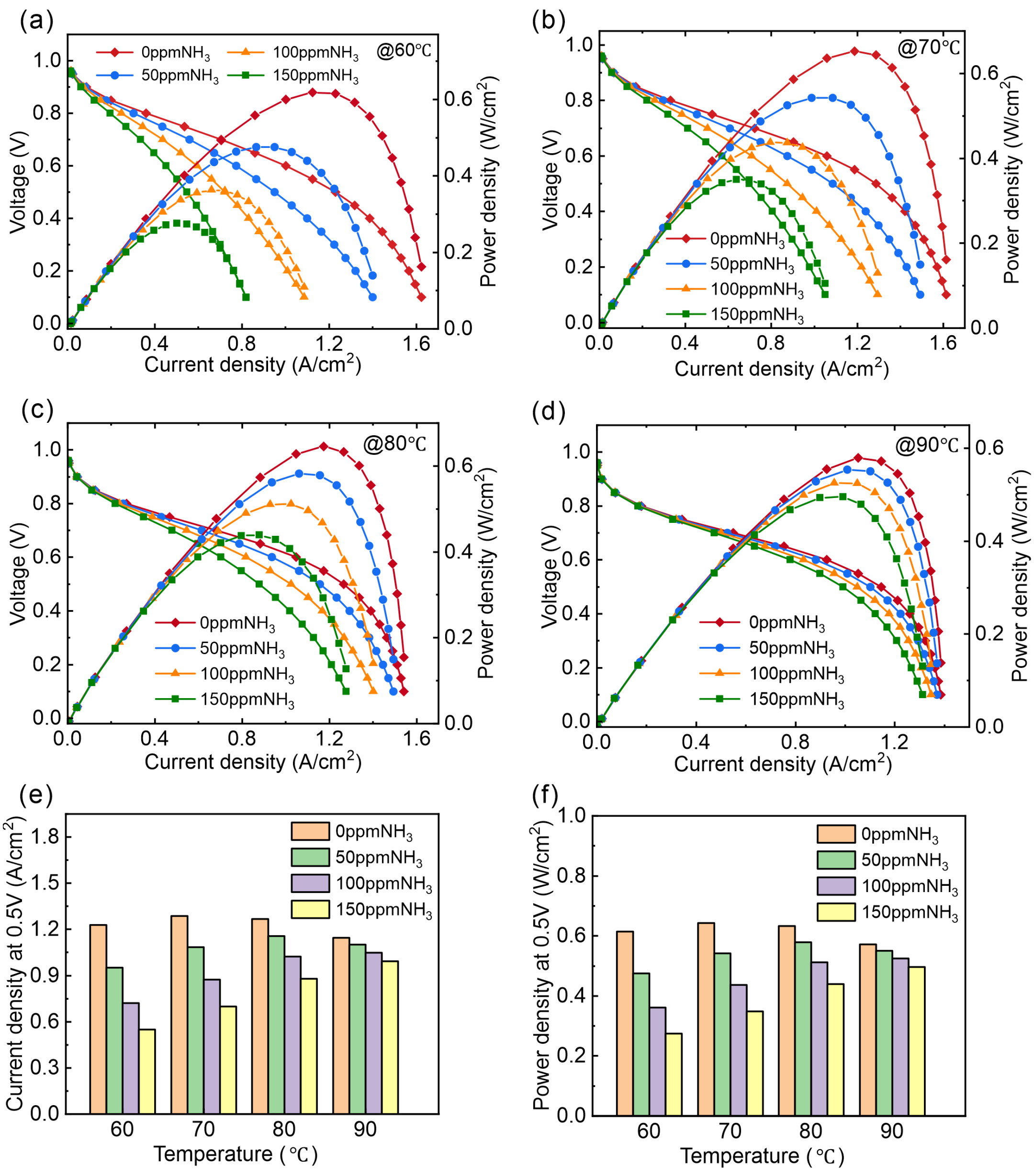

Figure 6. (a-d) Polarization and power-density curves of PEMFC at different operating temperatures and $NH_3$ concentrations: (a) 60 °C; (b) 70 °C; (c) 80 °C; (d) 90 °C. (e) Current density at 0.5 V. (f) Power density at 0.5 V.

## 3.3 Effect of membrane thickness

The membrane thickness critically influences the fuel cell performance. When PEMFCs undergo an electrochemical reaction, hydrogen is converted into protons and then passes through the PEM to the cathode to react with oxygen. Different membrane thicknesses have different resistance to proton transport. Therefore, it is important to study $NH_3$ poisoning at different membrane thicknesses. Here, we selected four common membrane thicknesses of 10, 25.4, 50.8, and 127 μm, and their polarization curves are shown in Figure 7. Taking the ammonia concentration of 50 ppm as an example, the cell performance decreases the least at the membrane thickness of 10 μm, and the maximum power density decreases by only 2.77%, while it decreases by 20.1% at the membrane thickness of 127 μm. This indicates that the reduction of membrane thickness is conducive to mitigating the poisoning effect of $NH_3$ on fuel cells. The reason is that reducing the membrane thickness can shorten the proton transport distance, improve the conduction efficiency, and counteract the influence caused by the combination of $NH_3$ and proton, thereby improving the cell performance. These results further demonstrate that a thinner membrane can enhance the tolerance to ammonia contamination. It is worth noting that although the reduction of membrane thickness is conducive to increasing the current density, it will also lead to a large difference in the current density distribution, which will lead to an uneven distribution of thermal and swelling stress on the MEA (see Section S3 in the Supplementary Material, including the distribution of current density fields in the membrane in Figure S4). Therefore, when selecting the membrane thickness of PEMFCs, it is necessary to comprehensively weigh the coupling effects of membrane thickness on proton conduction, current density distribution, ammonia resistance, etc., so as to find a suitable membrane thickness and achieve a balance between ammonia resistance and overall cell performance.

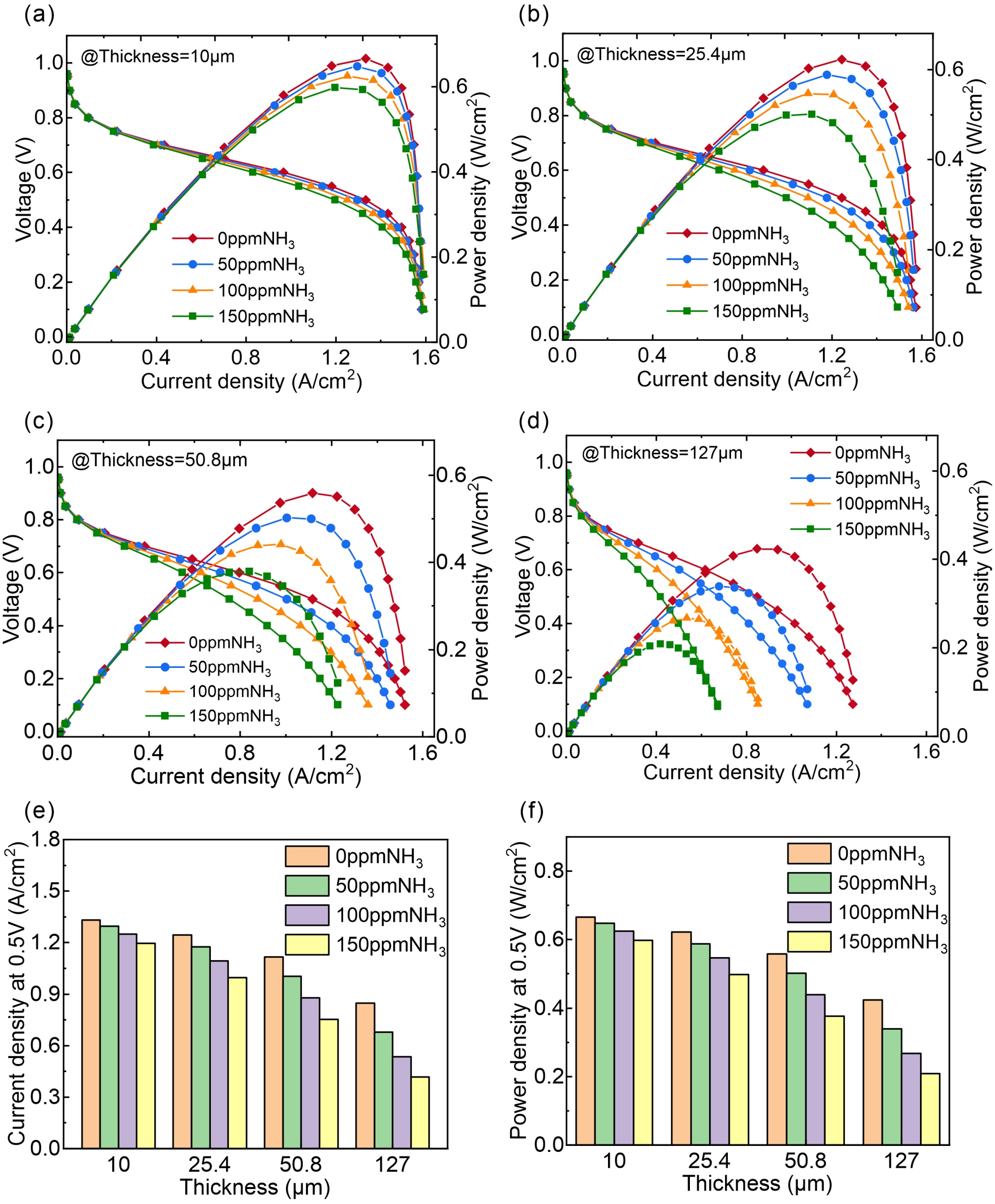

Figure 7. (a-d) Polarization and power-density curves of PEMFCs at different membrane thicknesses and $NH_3$ concentrations: (a) 10 μm; (b) 25.4 μm; (c) 50.8 μm; (d) 127 μm. (e) Current density at 0.5 V. (f) Power density at 0.5 V.

## 3.4 Influence of operating humidity

The operating humidity of PEMFCs serves as a pivotal factor influencing the cell performance. Gas humidification can improve the hydration level of PEM and promote proton transport within the membrane. However, excessive humidification may cause flooding in the PEMFC. Therefore, we study the influence of $NH_3$ impurity on cell performance under different operating humidities, as shown in Figure 8. Taking 150 ppm ammonia as an example, at 80%RH, the peak power density decreases by 32.7%, while at 60%RH, it decreases by 51.3%. This result indicates that $NH_3$ has less effect on proton conductivity at higher humidity, in alignment with the conclusion obtained by Hongsirikarn et al. [28,40]. This behavior can be explained by the hydration-dependent connectivity of proton-conducting pathways in PFSA ionomers. In PFSA-type ionomers, sulfonate-rich hydrophilic domains form water-containing clusters/nanochannels embedded in a hydrophobic matrix. At low RH (low hydration level $\lambda$), the sulfonate-rich hydrophilic domains are relatively small and poorly connected, leading to discontinuous water networks and limited proton-transport pathways; in this regime, partial neutralization by $NH_4^+$ imposes a larger effective transport penalty. As the relative humidity increases, the ionomer absorbs more water and swells, increasing both the size and connectivity of hydrophilic domains. This drives previously isolated ion clusters to become connected, forming continuous hydration pathways that facilitate proton conduction (enhanced Grotthuss hopping and lower effective activation barriers) [17,38]. Under $NH_3$ contamination, higher hydration further mitigates the effective transport penalty associated with partial site neutralization by improving ionic-domain connectivity and water-mediated screening/solvation, thereby alleviating the loss of membrane conductivity and cell performance.

The influence of $NH_3$ on fuel cells under varying relative humidity can be further verified from the proton conductivity distribution of the PEM and ACL. The increase in the operating humidity is conducive to improving the proton conductivity and mitigating the $NH_3$ poisoning, thereby enhancing the cell performance (see Section S4 in the Supplementary Material, including the proton conductivity in the membrane in Figure S5 and the proton conductivity and dissolved water content in the ACL in Figure S6). However, it should be noted that excessive humidification in the operation of actual cells may lead to liquid water accumulation, causing liquid-water flooding. Therefore, the combined effects of proton conductivity and water management should be considered in selecting the proper operating humidity.

The simulated trends are consistent with experimental observations reported in the literature, including polarization degradation, increased resistance/impedance contributions, and strong RH dependence under $NH_3$ exposure. Mechanistically, $NH_3$-induced site neutralization reduces the proton conductivity in the PEM and ACL ionomer, which increases the ohmic loss and redistributes the local current density; the resulting voltage loss is further modulated by RH and temperature through hydration and water-transport coupling (electroosmotic drag and back diffusion). From the perspective of commonly used experimental diagnostics, the predicted conductivity reduction corresponds to an increased ohmic contribution, which would typically be reflected by a higher high-frequency resistance (HFR) and a larger high-frequency intercept in electrochemical impedance spectroscopy (EIS). Moreover, the simulated current-density redistribution (under-rib vs under-channel and along-channel variations) can be examined using segmented-cell measurements or local current mapping.

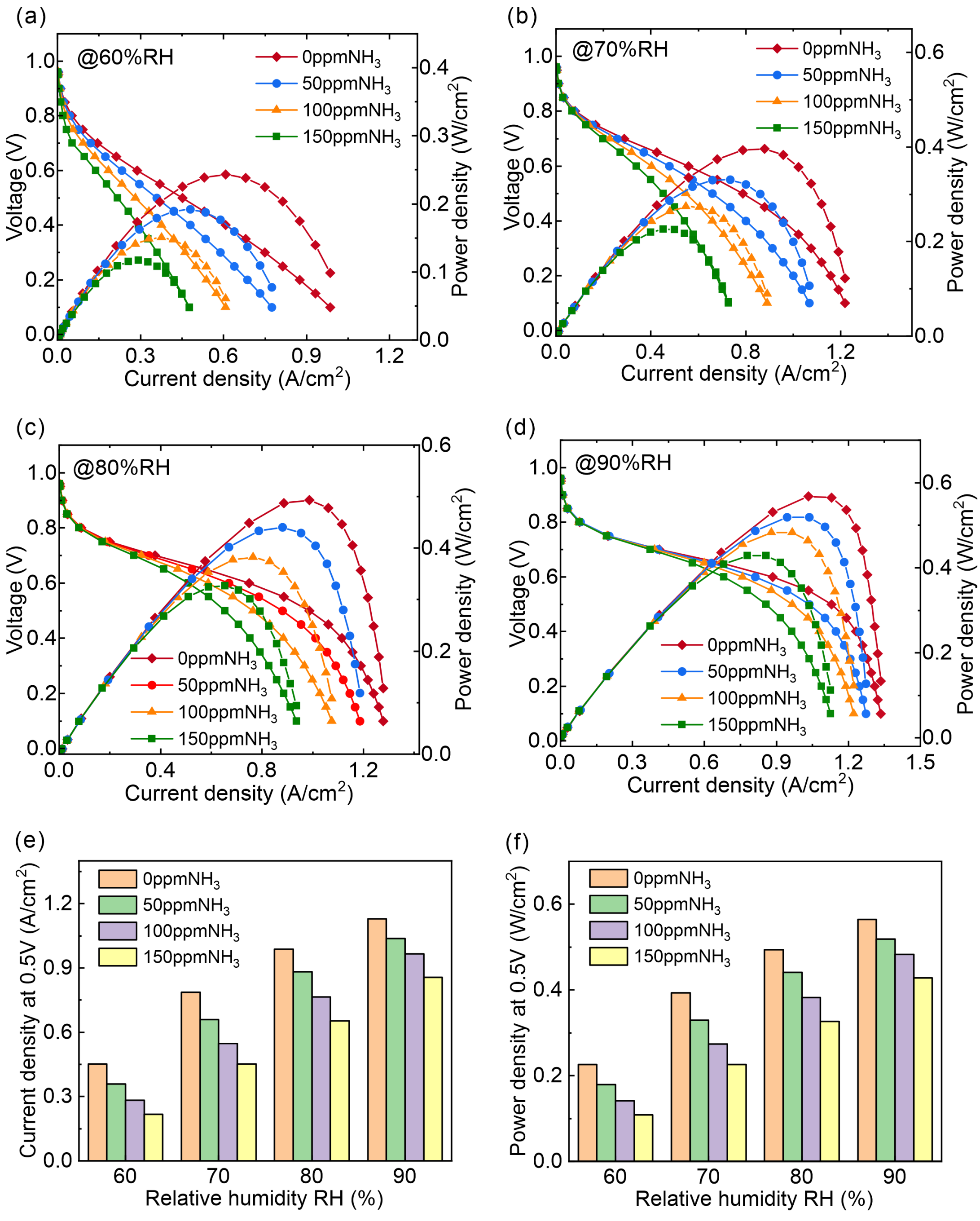

Figure 8. (a-d) Polarization and power-density curves of PEMFCs under different operating humidities and $NH_3$ concentrations: (a) 60%RH; (b) 70%RH; (c) 80%RH; (d) 90%RH. (e) Current density at 0.5 V. (f) Power density at 0.5 V.

The above parametric results can be used to provide practical guidance for evaluating $NH_3$ sensitivity, although they should not be interpreted as a universal optimization rule. The $NH_3$ tolerance limit depends on the MEA materials, flow-field design, operating protocol, durability requirement, and allowable performance loss of a specific application. Therefore, a useful way to define the tolerance concentration is to use a prescribed maximum-power-density loss $\Delta P_{\max}^{*} = \frac{P_{\max}(0) - P_{\max}(C_{NH3})}{P_{\max}(0)}$, where $P_{\max}(0)$ and $P_{\max}(C_{NH3})$ are the maximum power densities without and with $NH_3$ contamination, respectively. Within the investigated parameter range, the results suggest several trends that may help mitigate $NH_3$-induced performance loss. Increasing the operating temperature tends to reduce the predicted poisoning loss; at 150 ppm $NH_3$, the peak power-density loss decreases from 55.4% at 60 °C to 13.4% at 90 °C. However, this trend should be balanced against membrane dehydration, water balance, material stability, and thermal-management constraints. Increasing humidification also tends to improve tolerance to $NH_3$-induced ionomer conductivity loss. At 150 ppm $NH_3$, the peak power-density loss decreases from 51.3% at 60%RH to 32.7% at 80%RH. Nevertheless, excessive humidification may increase the risk of flooding, so the humidity condition should be selected by balancing ionomer hydration and water removal. The membrane-thickness results indicate that thinner membranes may reduce the proton-transport penalty caused by $NH_3/NH_4^+$-induced site neutralization. At 50 ppm $NH_3$, the maximum power-density loss is 2.77% for a 10 μm membrane and 20.1% for a 127 μm membrane. However, this trend should be interpreted together with gas crossover, mechanical durability, current-density uniformity, and long-term degradation considerations. Therefore, the present results provide quantitative sensitivity trends and a criterion for defining application-specific $NH_3$ tolerance.

# 4. Conclusions

In this study, a 3D numerical model of PEMFC with $NH_3$ impurities is established, and the transport process and cell performance under different $NH_3$ concentrations, operating temperatures, operating humidity, and membrane thickness are numerically studied. The underlying mechanism of ammonia on the transport process is analyzed from the perspectives of proton conductivity, current density, and dissolved water content. The results show that the presence of $NH_3$ impurity in hydrogen can significantly reduce the cell performance, mainly because $NH_3$ significantly reduces the proton conductivity of the PEM and the ACL. A higher $NH_3$ concentration leads to lower dissolved water content and lower proton conductivity of the PEM and the ACL, hence reducing the cell performance. When $NH_3$ concentration increases to 100 ppm, the peak power density decreases by 21.1%. The performance of cells containing $NH_3$ impurities can be enhanced by adjusting working conditions. A higher temperature can enhance the catalyst activity and reaction rate, and a higher operating humidity reduces the energy barrier of proton conductivity. Therefore, at higher temperatures and higher humidity, the $NH_3$ resistance of the fuel cell is improved. A thinner membrane can enhance the proton conduction rate within the membrane, thus improving the ammonia resistance of the cell, but at the same time, it will induce greater current density non-uniformity. The results are useful for the design and operation of PEMFCs and promote the utilization of ammonia-based hydrogen energy.